\documentstyle[prl,twocolumn,aps]{revtex}
\input{epsf.tex}
\newcommand{\bm}[1]{\mbox{\boldmath$#1$}}
\begin{document} \draft 
\twocolumn[
\title{Renormalization group analysis of the spin-gap phase
in the one-dimensional $t$-$J$ model}
\author{Masaaki Nakamura\cite{email1},
Kiyohide Nomura\cite{email2}, and Atsuhiro Kitazawa\cite{email3}}
\address{\cite{email1,email2,email3}
Department of Physics, Kyushu University, Fukuoka 812-81, Japan}
\address{\cite{email1,email3}
Department of Physics, Tokyo Institute of Technology,
Oh-Okayama, Meguro-ku, Tokyo 152, Japan}
\date{\today}\maketitle 
\begin{abstract}
\widetext\leftskip=0.10753\textwidth \rightskip\leftskip
 We study the spin-gap phase in the one-dimensional $t$-$J$ model,
 assuming that it is caused by the backward scattering process.
 Based on the renormalization group analysis and symmetry,
 we can determine the transition point between the
 Tomonaga-Luttinger liquid and the spin-gap phases,
 by the level crossing of the singlet and the triplet excitations.
 In contrast to the previous works, the obtained spin-gap region
 is unexpectedly large.
 We also check that the universality class of the transition belongs to the
 $k=1$ SU(2) Wess-Zumino-Witten model.
\end{abstract}
] \narrowtext
The existence of a gap in the spin excitation has been considered to be
a key to understand high-$T_c$ superconductivity.
This stimulated the study of one-dimensional (1D) electron systems
some years ago.
Recently, possibilities of superconductivity in quasi 1D systems
are suggested\cite{Dagotto-R},
and understanding of spin-gap phase in (quasi-)1D systems
increases the importance. Now, we reconsider this problem
from the 1D $t$-$J$ model which is the simplest but not fully understood.

The Hamiltonian of the 1D $t$-$J$ model is written as
\begin{eqnarray}
{\cal H}&=&-t\sum_{i\sigma}(c^{\dag}_{i\sigma} c_{i+1\sigma}
                   +c^{\dag}_{i+1\sigma} c_{i\sigma})\nonumber\\
 &&+J\sum_{i}(\hat{\bm{S}}_i\cdot\hat{\bm{S}}_{i+1}
  -\hat{n}_i \hat{n}_{i+1}/4),\label{eqn:t-J}
\end{eqnarray} 
in the subspace without double occupancy.
Generally, 1D electron systems belong to the universality class
of Tomonaga-Luttinger (TL) liquid\cite{Haldane,Solyom}
which is characterized by
gapless charge and spin excitations and
power-law decay of correlation functions.
The phase diagram of the 1D $t$-$J$ model is obtained
by Ogata {\it et al.}, using exact diagonalization\cite{Ogata-L-S-A}.
They found the enhancement of the superconducting correlation
($K_c > 1$)
and the phase separation ($K_c\rightarrow\infty$) for large $J/t$ region.
They also found a phase of singlet bound electron pairs
in the very low density region,
but could get no evidence for a spin-gap phase
by using a finite size scaling method at $1/3$ filling.
Hellberg and Mele studied this phase
by using a Jastrow-type variational wave function\cite{Hellberg-M}.
In their approach, the variational parameter $\nu$
is related with $K_c$ as $K_c=1/(2\nu + 1)$.
They found that there exists a finite region where the optimized parameter
takes constant value $\nu=-1/2$ between TL phase and phase-separated state,
and they interpreted the region as the spin-gap phase.
Other variational wave function is proposed by Chen and Lee\cite{Chen-L}.

However, these authors did not discuss the detailed mechanism of the
spin gap generation.
One candidate of the spin gap generation mechanism is due to the
attractive backward scattering
(scattering between electrons with the opposite momentum ($k_F,-k_F$)
and spin)\cite{Manyhard-S,Solyom}.
In this case, the universality class of the transition
is the $k=1$ SU(2) Wess-Zumino-Witten (WZW) model\cite{Affleck-G-S-Z}.
On the basis of this assumption,
we determine the transition point
with the singlet-triplet level crossing method
\cite{Ziman-S,Affleck-G-S-Z,Okamoto-N}
and we obtain the phase diagram (FIG.\ref{fig:PD-NNK}).
Then we will verify the consistency of our method,
considering the ratio of the logarithmic correction term.

In general, the low-energy behavior of a 1D electron system
is described by the U(1) Gaussian model (charge part)
and the SU(2) sine-Gordon model (spin part)\cite{Solyom,Schulz},
\begin{equation}
 {\cal H}={\cal H}_c + {\cal H}_s
  + \frac{2 g_1}{(2\pi\alpha)^2}\int dx \cos(\sqrt{8}\phi_s)
  \label{eqn:effHam}.
\end{equation}
Here $\alpha$ is a short-distance cutoff, $g_1$ is the backward
scattering amplitude, and for $\nu=c,s$
\begin{equation}
 {\cal H}_{\nu}=\frac{1}{2\pi}\int dx
 \left[v_\nu K_\nu(\pi\Pi_\nu)^2+\frac{v_\nu}{K_\nu}
 \left(\frac{\partial\phi_\nu}{\partial x}\right)^2\right]
  \label{eqn:Gaussian},
\end{equation}
where $\Pi_\nu$ is the momentum density conjugate to $\phi_\nu$,
$[\phi_\nu(x),\Pi_\nu(x')]=i\delta(x-x')$,
$K_\nu$ is the Gaussian coupling,
and $v_c$ and $v_s$ are charge and spin velocities, respectively.
The primary field of this model is
$\exp i\sqrt{2}(m_{\nu}\phi_{\nu}+n_{\nu}\theta_{\nu})$,
where the dual field is defined as $\partial_x\theta_{\nu}=\pi\Pi_{\nu}$.
In TL phase ($g_1>0$), the parameters $K_s$ and $g_1$ will be renormalized as
$K^{*}_s=1$ and $g_1^{*}=0$, reflecting the SU(2) symmetry.

First, let us consider the case without renormalization, $g_1=0$.
The finite size correction of the energy and the momentum
of (\ref{eqn:Gaussian}) are described by the conformal field theory
(CFT)\cite{Blote-C-N,Cardy84} with $c=1$,
where the central charge $c$ characterizes
the universality class of the model.
For the $t$-$J$ model, $c=1$ as shown rigorously
at $J/t=2$\cite{Kawakami-Yang90} and numerically\cite{Ogata-L-S-A}.
The combined use of the CFT and the Bethe ansatz result
gives a description of the 1D electron systems
\cite{Woynarovich,Frahm-K,Bares-B-O,Kawakami-Yang90}.
The ground state energy of the system
under periodic boundary conditions is given by
\begin{equation}
 E_0(L)=L\epsilon_0-\frac{\pi(v_c+v_s)}{6L}c,
\end{equation}
where $L$ is the system size.
The excitation energy and momentum are related with exponents as
\begin{eqnarray}
 E-E_0&=&\frac{2\pi v_c}{L}x_c+\frac{2\pi v_s}{L}x_s,\label{eqn:energy}\\
 P-P_0&=&\frac{2 \pi}{L}(s_c+s_s)+4k_F D_c + 2k_F D_s
      \label{eqn:momentum},
\end{eqnarray}
where $k_F=\pi N/2L$ with electron number $N$,
and the scaling dimensions and the conformal spins are defined by
$x_\nu=\Delta_\nu^+ + \Delta_\nu^-,
 s_\nu=\Delta_\nu^+ - \Delta_\nu^-$, respectively,
with the conformal weights,
\begin{equation}
\Delta_{\nu}^{\pm}=\frac{1}{2}
    \left(\sqrt{\frac{K_{\nu}}{2}}m_{\nu}
     \pm\frac{n_{\nu}}{\sqrt{2K_{\nu}}}\right)^2+N_{\nu}^{\pm}.
\end{equation}
The variables $m_{\nu}$ and $n_{\nu}$ are related
with electron quantum numbers as
$m_c=2D_c+D_s,n_c=\Delta N_c/2,m_s=D_s,n_s=\Delta N_s-\Delta N_c/2$.
Here $\Delta N_c$ is the change of
the total number of electrons, and $\Delta N_s$ is
the change of the number of down spins.
$D_{c}$ ($D_{s}$) denotes the number of particles moved
from the left charge (spin) Fermi point to the right one.
$N_{c}^{\pm}$ ($N_{s}^{\pm}$) is characterized by simple particle-hole
excitations near right or left charge (spin) Fermi points.

These quantum numbers are restricted by the selection rule
under periodic boundary conditions\cite{Woynarovich}
\begin{mathletters}
\begin{eqnarray}
 D_c&=&\frac{\Delta N_c+\Delta N_s}{2}
\ (\mbox{mod}\ 1),\label{eqn:dDa}\\
 D_s&=&\frac{\Delta N_c}{2}\hspace{1.3cm} (\mbox{mod}\ 1).\label{eqn:dDb}
\end{eqnarray}
\end{mathletters}
In the case of twisted boundary conditions
$c^{\dag}_{j+L,\sigma}=e^{i\Phi}c^{\dag}_{j\sigma}$
which is equivalent to the system
where the flux $\Phi$ penetrates the ring\cite{Kohn},
$D_c$ is modified as $D_c+\Phi/2\pi$.
For the ground state $E_0$,
we choose periodic boundary conditions ($\Phi=0$) for $N=4m+2$ electrons
and antiperiodic boundary conditions ($\Phi=\pi$) for $N=4m$ electrons
with an integer $m$.
Changing the boundary conditions, the ground state becomes always
singlet with zero momentum ($P_0=0$)\cite{Ogata&Shiba,Ogata-L-S-A}.

In order to eliminate the contribution of the charge part, and extract
the singlet and the triplet excitation in the spin part ($x_s=1/2$),
we turn our attention on following states:
$(\Delta N_c, \Delta N_s, D_c, D_s)
=(0,\pm 1,0,0),(0,0,\mp 1/2,\pm 1)$ under twisted boundary conditions
($\Phi=\pi$ for $N=4m+2$, $\Phi=0$ for $N=4m$).
We can identify these excitation spectra
by using (\ref{eqn:energy}) and (\ref{eqn:momentum}),
but the momentum $P$ and the wave number $p$ are not always identical.
There is a relation $P=p-\Phi N/L$ between them\cite{momentum_shift}.

Next, we consider the renormalization ($g_1\neq 0$).
By the change of the cut off $\alpha\rightarrow e^{dl}\alpha$,
the coupling constant $g_1$ and $K_s$ are renormalized as\cite{Kosterlitz}
\begin{mathletters}
\begin{eqnarray}
 \frac{dy_0(l)}{dl}&=&-y_1^{\ 2}(l),\\
 \frac{dy_1(l)}{dl}&=&-y_0(l) y_1(l),
\end{eqnarray}
\end{mathletters} 
where $y_1(l)=g_1/\pi v_s, K_s=1+y_0(l)/2$.
For the SU(2) symmetric case $y_{0}(l)=y_{1}(l)$, and $y_0(l)>0$,
the scaling dimensions of the operators for
singlet and triplet excitations
$\sqrt{2}\cos\sqrt{2}\phi_{s}$ ($x_{ss}$), and
$\sqrt{2}\sin\sqrt{2}\phi_{s}, 
\exp(\mp i\sqrt{2}\theta_{s})$ ($x_{st}$)
split logarithmically by the marginally irrelevant coupling as
\cite{Giamarchi-S}
\begin{mathletters}
\begin{eqnarray}
x_{ss}&=&\frac{1}{2}+\frac{3}{4}\frac{y_0}{y_0\log L+1},\\
x_{st}&=&\frac{1}{2}-\frac{1}{4}\frac{y_0}{y_0\log L+1},
\end{eqnarray}
 \label{eqn:sclngdim}
\end{mathletters}
where $y_0$ is the bare coupling, and we have set $l=\log L$.
This result is equivalent with that of the $k=1$ SU(2) WZW model
\cite{Affleck-G-S-Z}.
Note that the ratio of the logarithmic corrections are
given by Clebsch-Gordan coefficients.
When $y_0<0$, $y_0(l)$ is renormalized to $y_0(l)\rightarrow -\infty$,
and there appears spin gap.
At the critical point ($y_0=0$),
there are no logarithmic corrections in the excitation gaps.
The physical meaning of this point is
that the backward scattering coupling changes from
repulsive to attractive.
And the SU(2) symmetry is enhanced at the critical point
to the chiral SU(2)$\times$SU(2) symmetry\cite{Affleck-G-S-Z},
since the spin degrees of freedom of the right and the left Fermi points
become independent.
Therefore, the critical point is obtained from
the intersection of the singlet and the triplet excitation spectra
\cite{Ziman-S,Affleck-G-S-Z,Okamoto-N}.
Using this method, we can determine the critical point
with high precision\cite{Okamoto-N},
since the remaining correction is only
$x_s=4$ irrelevant fields\cite{Cardy86,Reinicke}.

\begin{figure}
\epsfxsize=2.8in \leavevmode \epsfbox{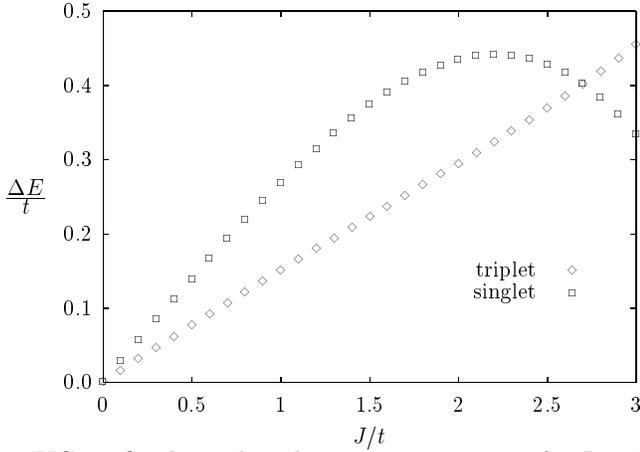}
\caption{Singlet and triplet excitation energies
 for $L=16$ system at $n=1/2$.}
\label{fig:EXAMPLE}
\end{figure}

\begin{figure}
\epsfxsize=2.8in \leavevmode \epsfbox{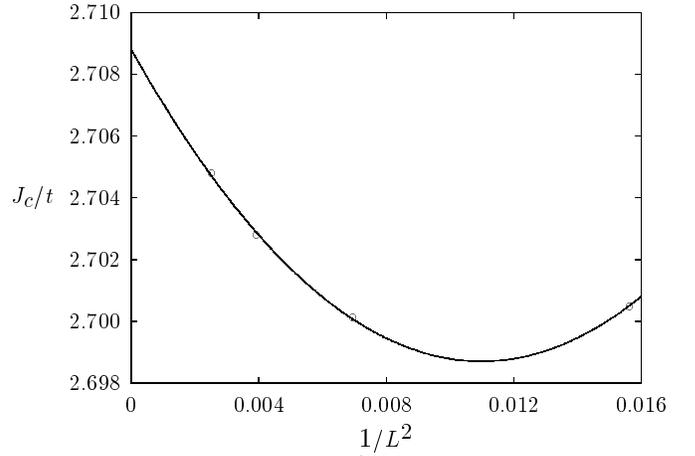}   
\caption{Size dependence of $J_c/t$ determined by the intersections
 of the excitation spectra for $L=8,12,16,20$ systems at $n=1/2$.
 These points are fitted by the form $A+B/L^2+C/L^4$.}
\label{fig:SIZE-QF}
\end{figure}

Here we analyze the numerical results for the $t$-$J$ model
(\ref{eqn:t-J}) with the above explained method.
We diagonalize $L=8$-$30$ systems
by the use of Lanczos and Householder method.
An example of data ($L=16,n\equiv N/L=1/2$) is shown in FIG.\ref{fig:EXAMPLE}.
Since the critical point is almost independent
of the system size as is shown in FIG.\ref{fig:SIZE-QF},
the phase diagram can be constructed without extrapolation.
Our result is similar to the Hellberg and Mele's in the low density region,
but the spin-gap phase spreads extensively toward the high density region.
We are not able to answer whether the spin gap survives
in the $n\rightarrow 1$ limit or not,
because the numerical results become unstable in the high density region
where the phase boundary is close to the phase-separated state.
In TL phase, singlet and triplet superconducting correlations (SS, TS)
have the same critical exponent $1/K_c+1$\cite{Solyom},
while with a spin gap, TS decays exponentially and SS is enhanced as $1/K_c$,
so that SS is dominant in the spin-gap region.

\begin{figure}
\epsfxsize=2.8in \leavevmode \epsfbox{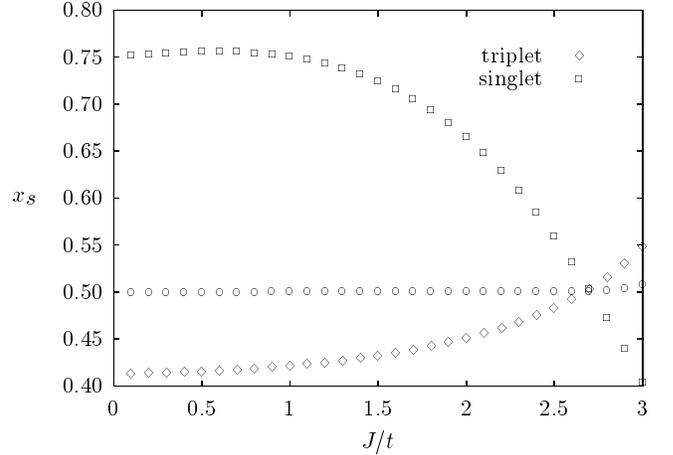}
\caption{Extrapolated value of $(x_{ss}+3x_{st})/4$
 and the scaling dimensions for the singlet ($x_{ss}$)
 and the triplet ($x_{st}$) excitations for $L=16$ system at $n=1/2$.}
\label{fig:RATIO}
\end{figure}

\begin{figure}
\epsfxsize=2.8in \leavevmode \epsfbox{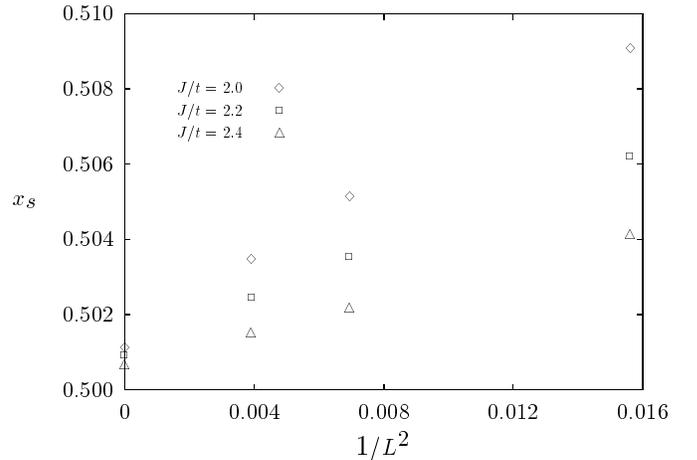}
\caption{Size dependence of the averaged scaling dimension
  $(x_{ss}+3x_{st})/4$ at $n=1/2$.}
\label{fig:size_xs}
\end{figure}

In order to check the consistency of our argument,
we calculate the ratios of the logarithmic corrections
and scaling dimensions for the singlet and the triplet excitations
from (\ref{eqn:energy}) and (\ref{eqn:sclngdim}).
Here the spin wave velocity is given by\cite{velocity}
\begin{equation}
 v_s=\lim_{L\rightarrow\infty}
  \frac{E(L,N,S=1,P=2\pi/L)-E_0(L,N)}{2\pi/L},
\end{equation}
which is extrapolated by the function
$v_s(L) = v_s(\infty) + A/L^2 + B/L^4$.
These corrections are explained by the irrelevant fields.
The average of the renormalized scaling dimension $(x_{ss}+3x_{st})/4$,
eliminating logarithmic corrections,
and its finite size effect are shown in FIG.\ref{fig:RATIO}
and FIG.\ref{fig:size_xs}, respectively.
The extrapolated data become $1/2$ with error less than 0.2 \%.

Finally, we discuss the reason why the previous studies
have estimated the spin gap region very narrower than
the real one.
From the two-loop renormalization group equation
of the $k=1$ SU(2) WZW model\cite{Amit-G-G,Destri,Nomura}
\begin{equation}
 \frac{dy_0(l)}{dl}=-y_0^{\ 2}(l)-\frac{1}{2}y_0^{\ 3}(l),
\end{equation}
the spin gap $\Delta E$ grows singularly as
\begin{equation}
 \Delta E\propto\sqrt{J-J_c}\exp(-\mbox{Const.}/(J-J_c)),
  \label{eqn:gap}
\end{equation}
where $y_0\propto J_c-J$,
therefore it is very difficult to find the critical point using conventional
finite size scaling method.
Note that (\ref{eqn:gap}) is the same asymptotic behavior
as the spin gap of the negative $U$ Hubbard model at half-filling,
which can be obtained from the charge gap at positive $U$\cite{Ovchinikov},
and the transformation between the charge and the spin
degrees of freedoms\cite{Shiba}.

In conclusion, we studied the spin-gap phase in the 1D $t$-$J$ model,
considering the backward scattering effect in the TL liquid
by the renormalization group analysis.
Using the twisted boundary conditions,
we can extract the spin excitation spectra,
and determine the critical point as in spin systems.
The phase boundary is determined by the point where the
backward scattering becomes repulsive to attractive.
The spin-gap phase obtained in this way is unexpectedly large,
and the consistency of the argument is also checked.
This method can be applied to other models in 1D electron systems,
if the SU(2) symmetry is assured.

This work is partially supported by Grant-in-Aid for
Scientific Research (C) No. 09740308 from the Ministry of Education,
Science and Culture, Japan.
A.K. is supported by JSPS Research Fellowships for Young Scientists.
The computation in this work was done
using the facilities of the Supercomputer Center,
Institute for Solid State Physics, University of Tokyo.

\begin{figure}
 \epsfxsize=3.3in \leavevmode \epsfbox{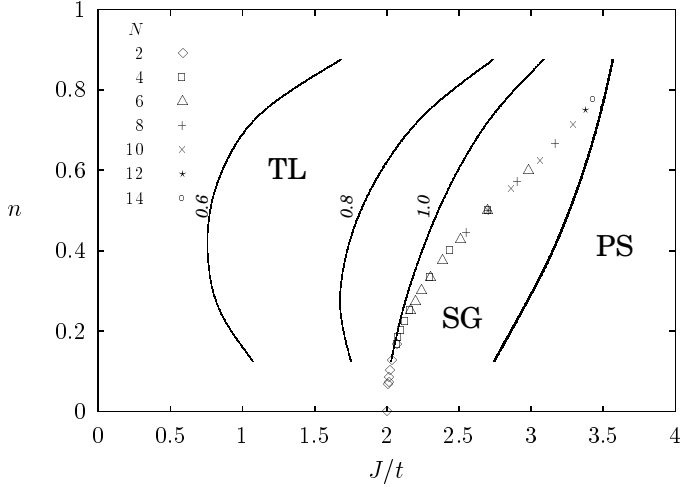}
\caption{Phase diagram of the 1D $t$-$J$ model
  (TL: TL phase, SG: spin-gap phase, PS: phase-separated state).
 In the spin-gap phase where the backward scattering is attractive,
 the singlet excitation becomes lower than the triplet
 (see FIG.\protect{\ref{fig:EXAMPLE}},\protect{\ref{fig:RATIO}}).
 The contour lines of $K_c$ are calculated by the data of $L=16$ system
 \protect{\cite{contour}}.}
\label{fig:PD-NNK}
\end{figure}

\begin{references}
\bibitem{velocity}
 The quantum numbers correspond to the excitation for the spin velocity are
 $(\Delta N_c,|\Delta N_s|,D_c,D_s)=(0, 1,\pm1/2,\mp 1)$
 or $(N_s^+,N_s^-)=(1,0),(0,1)$.
 All these states have momentum $P=\pm 2\pi/L$,
 and form the SU(2) triplets.
\end{references}
\end{document}